\documentstyle[12pt,epsf]{article}
\begin{document}
\begin{titlepage}
\pagestyle{empty}
\baselineskip=21pt
\rightline{Alberta Thy-41-94}
\rightline{December 1994}
\vskip .2in
\begin{center}
{\large{\bf Fermi Ball Detection}}
\end{center}
\vskip .1in
\begin{center}
Alick L. Macpherson and James L. Pinfold

{\it Department of Physics, University of Alberta}

{\it  Edmonton, Alberta, Canada T6G 2J1}

\vskip .2in

\end{center}

\begin{abstract}

The detectability of charged SLAC-bag type structures is considered. These
objects, known as Fermi balls, arise from the spontaneous symmetry
breaking of a biased discrete symmetry in the early universe. Two classes
of experimental searches are discussed. Since Fermi balls in the
theoretically favoured mass range are absorbed by the atmosphere, direct
experimental searches are confined to space-based facilities. Simple
spectrometer and time of flight analysis give a quantitative estimate of
Fermi ball mass up to a limit set by the system's tracking resolution. For
the ASTROMAG facility, with a tracking resolution of 50 $\mu$m the upper
bound on detectable Fermi ball masses is of order $10^{15}$ GeV/c$^{2}$.
Charged tracks with sagitta smaller than this resolution would provide
evidence in favour of Fermi balls, but only give a lower bound on the
discrete symmetry breaking scale $\varphi_{0}$. The second class of
experimental search proposed relies on the detection of bound Fermi ball
states that have been concentrated in terrestrial materials such as oceanic
sediment.
\end{abstract}

\baselineskip=18pt
\end{titlepage}
\baselineskip=18pt

An analysis of biased discrete symmetry breaking in the early universe has
indicated the possibility of production of composite particles called
Fermi balls\cite{FB}. These Fermi balls are formed out of collapsing
fermion-populated domain walls that are generated as the result of
spontaneous symmetry breaking of a biased discrete symmetry associated
with a real scalar field. A strong Yukawa coupling of generic fermions to
this real scalar field insures that fermions are swept up by, and stay
within, the domain walls as they collapse upon themselves, thereby forming
finite sized false vacuum bags enclosed by a fermion populated domain wall
skin.

These false vacuum bags collapse and fragment until the soliton nature of
the bag structure arrests the collapse. Conceptually, this halting of the
collapse is a result of the Fermi gas pressure of the domain wall fermions
balancing the false vacuum volume pressure and domain wall surface
tension. In the bag model description of the Fermi ball this may occur
when the thin domain wall approximation breaks down. The structure that
emerges is one of numerous composite particles (Fermi balls) each composed
of massless fermions contained in a supermassive SLAC bag like construct
with a radius ($ \mbox{in GeV}^{-1}$)
\begin{equation}
 {R}_{FB} \sim {2 \over {\varphi}_{0}}
\label{rad}
\end{equation}
and a mass of approximately $100 \varphi_{0}$ GeV/$c^{2}$ where, $
\varphi_{0}$ (in GeV) is the symmetry breaking scale. Hence, the Fermi ball
mass is dependent on the discrete symmetry breaking scale parameter
$\varphi_{0}$.

The spontaneous symmetry breaking of a biased discrete symmetry is not in
itself sufficient to produce cosmologically stable Fermi balls. Such
objects can only exist if there exists a net fermion antifermion
asymmetry. As the domain wall confinement of fermions prevents fermion
number freeze out, a Fermi ball would be completely deflated by
fermion pair annihilations if there was a fermion antifermion symmetry.
Assuming a fermion antifermion asymmetry, these cosmologically stable
Fermi balls can carry a standard model gauge charge which depends on the
fermion content of the individual Fermi ball.

Conservative constraints on the neutral Fermi ball mass and cross section
have already been given in \cite{FB}. We focus on the detection of Fermi
balls with overall standard model gauge charges, and for simplicity
consider the case of an electric charge equal to the sum of the charges of
the Fermi ball fermion population. A specific Fermi ball charge prediction
can only be the result of a detailed study of fermion-antifermion
asymmetries just prior to discrete symmetry breaking, subsequent domain
wall formation, and fermion evaporation and reabsorption. In order to
minimise assumptions as to extent of fermion antifermion asymmetry in the
early universe we allow for a Fermi ball charge ranging from -Ne to +Ne,
where N is the number of massless fermions contained in the Fermi ball --
N$\sim 50$ \cite{FB}, independent of the breaking scale. If more than one
fermion type is present, such a mixture of fermion species would only
serve to reduce the Fermi ball charge from the maximum allowable charge
($\pm$ Ne).

Fermi ball production is the result of the collapse and fragmentation of
false vacuum bubbles encased in fermion populated domain walls into
massive remnants. Consequently, assuming no special acceleration
mechanisms are operating, one would expect the typical Fermi ball velocity
to be of the order of the average galactic velocity $v \sim 250$ km/s , or
less.  For a Fermi ball with a typical velocity of order $10^{-3}$c, the
quantitative estimate of the mass required for a maximally charged Fermi
ball to penetrate the atmosphere depends on the sign of the charge.
Assuming positively charged Fermi balls generate a completely neutralising
electron cloud as they pass through the atmosphere, the Fermi ball is
analogous to a nuclearite, and has an energy loss per path length given by
De Rujula and Glashow \cite{nuclearite}:

\begin{equation}
\frac{dE}{dx} = -A\rho v^2
\end{equation}
where $A$ is the effective cross-sectional area of the nuclearite, $v$ is
its velocity, and $\rho$ is the density of the medium. Thus $ v$,
decreases
exponentially with distance D, according to:
\begin{equation}
 v(D) = v(0)e^{-(\frac{A}{M}\int^{D}_{0}\rho dx)}
\end{equation}
where M is the mass of the nuclearite. Taking the column density of the
atmosphere to be 1013 g/cm$^2$, the mass required for the positively
charged Fermi ball to penetrate the atmosphere and retain a cosmic
velocity ($\beta = 10^{-3}$) is of order $9 \times 10^{9}$ GeV (i.e.
$\varphi_{0} \approx 10^{8}$ GeV). Alternatively, negatively charged Fermi
balls suffer energy loss due primarily to electromagnetic interactions
with atomic electrons, and for such low velocity objects the energy loss
calculation is analogous to that for a charged heavy ion undergoing only
electromagnetic interactions. An approximate form of the energy loss per
path length has been given by Lindhard \cite{Lindhard}, which assigns no
specific structure to the projectile and treats the surrounding atomic
electrons as an electron gas of constant density. Lindhard's model assumes
the projectile forms no neutralising cloud, and so for the energy loss
calculation, the Fermi ball acts like an ion of atomic number $Z_{1} =
|Q|$, where $Q$ is the bare Fermi ball charge. The energy loss per path
length for such a slow moving negatively charged Fermi ball is then
estimated by:
\begin{equation}
{dE \over dx} =- {2m_{e}^{2} Z_{1}^{2} e^{4} v \over 3 \pi \hbar^{3}}
 (\log {137 v_{F} \over c} + \log \pi - 1 + {2 c \over 137 \pi v_{F}} ).
\end{equation}
where, for a typical detecting medium, the ambient electron velocity is
the Fermi velocity $v_{F}$, which is of order the Bohr velocity, $v_{0}=
{e^{2} \over \hbar } \approx 2.2 \times 10^{8}$ cm/s $= 7.3 \times 10^{-3}
c$. The mass of a negatively charged Fermi ball required to penetrate the
atmosphere and retain a velocity between $10^{-5}$c and $10^{-3}$c is
obtained by evaluation of the mean range $R= \int {dx \over dE} dE$. For a
maximally charged Fermi ball this mass is $10^{15}$ GeV or greater.

The experimental searches considered in this work are divided into two
categories: space-based and terrestrial, detection experiments. The choice
of two classes of experiment is determined by the fact that unless the
charged Fermi balls are extremely heavy, they will range out in the
atmosphere.  Maximum sensitivity for active searches is obtained using
space based experiments, as they offer the possibility of an experimental
search over the full range of $\varphi_{0}$. These space based experiments
need only be simple spectrometers, which when coupled with independent
time of flight and charge measurements, allow determination of charged
particle masses. As the experiment is space based, no neutralising cloud
is expected to form around the incident Fermi balls, and so the experiment
is sensitive to the bare charge (of either sign).

\begin{figure}[t]
\includegraphics{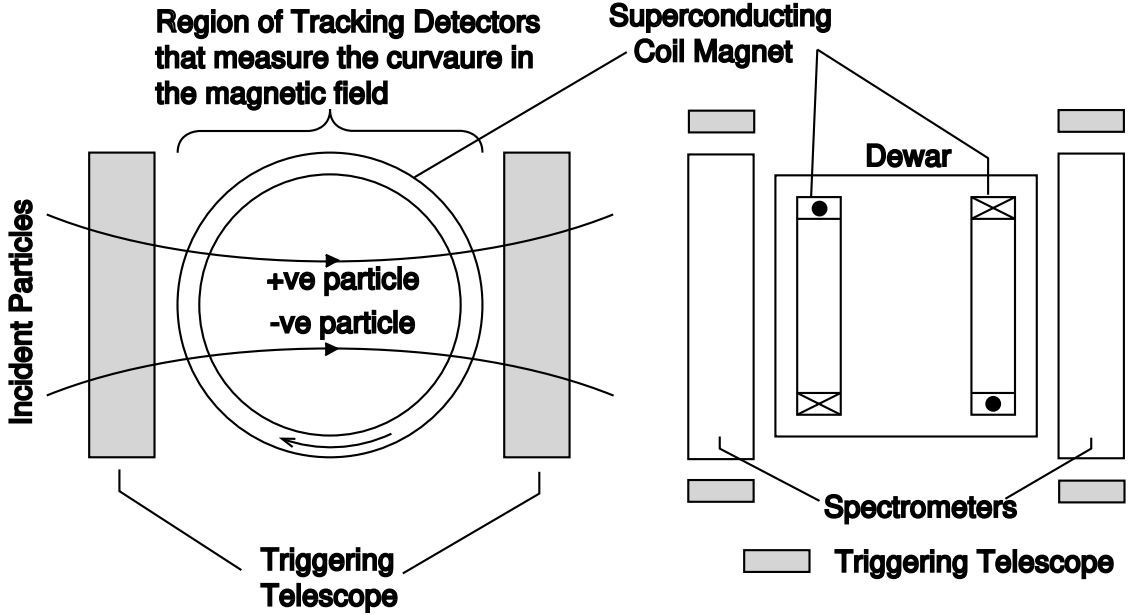}
\vspace{8.2cm}

\caption[]{Conceptual technique and layout of the ASTROMAG facility}


\end{figure}

Suitable experimental facilities, ASTROMAG \cite{Astromag} and WiZard
\cite{Wizard}, have been proposed. The conceptual layout of ASTROMAG is
that of a magnetic analyzer, triggering telescope and a data acquisition
system shown schematically in Figure 1. One proposed design for the
ASTROMAG facility has a thin superconducting solenoid with coil diameter
2m, a central magnetic field of $\sim$1.3 Tesla, and a tracking system
resolution of $50 {\mu}m$.  Identification of charged Fermi balls is
performed by measurement of the sagitta of a charge particle track in the
magnetic field. Negatively charged Fermi balls would be particularly
distinctive, especially if the magnitude of the Fermi ball charge is
maximal. Combined with the independent measurements time of flight
information and charge of the particle, the mass can easily be determined.
For charged Fermi balls, the signature that is expected is that of a
superheavy particle with a velocity of the order of $10^{-3}$c and a
maximum charge around 50e that can be positive or negative. As the Fermi
ball is expected to be extremely heavy the sagitta of its track could be
less than the resolution of the spectrometer. In this case the charged
Fermi ball signature would be that of a non-relativistic charged particle,
with positive or negative charge, that produces a track with no measurable
curvature. The occurrence of such tracks implies the presence of Fermi
ball candidates, but would only allow a lower bound on the discrete
symmetry breaking scale, $\varphi_{0}$ (as the Fermi ball mass
$M_{FB}\approx 100 \varphi_{0}$ GeV/c$^{2}$).

The mass range to which the spectrometer is sensitive is explored by
plotting the contours of the maximum value of the measurable mass as a
function of charge, assuming a uniform magnetic field and a Fermi ball
velocity of $10^{-3}$c.  Contours which are determined by setting the
sagitta to the value of the tracking system resolution give an upper bound
on the Fermi ball mass range to which the experiment is sensitive. Several
such contours are plotted in Figure 2. It can be seen that for an atomic
number of 26 and a tracking system resolution of $50 \mu$m, the maximum
identifiable mass by the method of sagitta evaluation is of the order of $
\sim 10^{8}$ Gev/c$^{2}$ -- this corresponds to only a moderate coverage
of the breaking scale parameter space : $\varphi_{0} < 10^{6}$.  By
comparison, the momentum required by an iron atom to produce a track with
a sagitta of $50 \mu$m, equal to the tracking system resolution, is $ 3
\times 10^{13}$ Gev/c -- far greater than the value of $0.052$ GeV/c for
an iron atom moving at $10^{-3}$c. This example illustrates that when
timing and sagitta information are combined the possibility of
misidentification of Fermi balls with heavy ions is effectively ruled out.

\begin{figure}[t]
\includegraphics{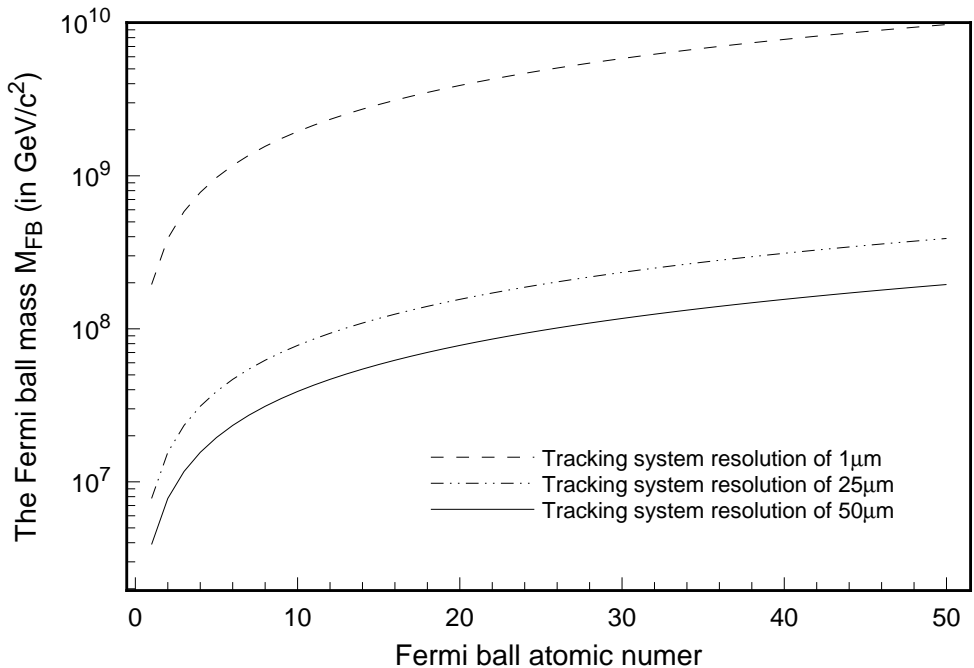}
\vspace{9cm}

\caption[]{Contour plots of the maximum value of measurable mass, as a
function of charge, obtained by setting the sagitta equal to various
values of the tracking system resolution. The expected resolution for
ASTROMAG is of the order of 50$\mu$m}


\end{figure}

Any direct Earth based search for charged Fermi balls is only directly
sensitive to the extreme end of the $\varphi_{0}$ parameter space.
However, there is the possibility Fermi balls that have been stopped in
the atmosphere are detectable in passive terrestrial experiments. For
example, a positively charged Fermi ball will acquire a neutralizing cloud
of electrons as it slows down in the Earth's atmosphere. These heavy
stopped Fermi balls, cloaked with a neutralizing charge, would fall to
earth and into the oceans. Fermi balls are characteristically expected to
be multiply charged, with a mass $100 \varphi_{0}$ GeV/$c^{2}$ and a
radius which varies as the reciprocal of $\varphi_{0}$, being $\sim$1
Fermi for $\varphi_{0}$= 1 GeV.

As characteristic breaking scales are substantially greater than 1 GeV we
can regard the Fermi ball and its accompanying electrons as a superheavy
atom with a maximum $Z$ of about 50e, where the radius of the Fermi ball
is typically substantially smaller that 1 Fermi. One possible technique
for detecting such an object over most of the possible mass range is time
of flight mass spectrometry \cite{TOF}. In this approach the sample
believed to contain Fermi balls, such as ocean sediment, is vapourized
with a laser beam and then fully or partially ionized. The time of flight
to a microchannel plate, or other suitable charged particle detector, of
the fully or partially ionized Fermi ball is then measured. Although, this
method does not directly detect the incoming Fermi ball its has the
advantage that the possible Fermi ball population in the terrestrial
material under test has presumably accumulated over billions of years.

\noindent{ {\bf Acknowledgements} } \\ \noindent We would
like to thank B. A. Campbell, F. C. Khana and G. Roy for useful
discussions.

This work was supported in part by the Natural Sciences and Engineering
Research Council of Canada.


\begin{thebibliography}{99}

\bibitem{FB} A. L. Macpherson and B. A. Campbell, University of Alberta
preprint, Alberta Thy-01-94 (1994). hep-ph 9408387

\bibitem{nuclearite} A. De Rujula and S. L. Glashow, Nature {\bf 312}, 734
(1984).

\bibitem{Lindhard} J. Lindhard, Mat. Fys. Medd. Dan. Vid. Selsk. {\bf 28},
No. 8 (1954).

\bibitem{Astromag} G. Auriemma, Nucl. Instr. and Meth. {\bf A263}, 243
(1988); A Yamamoto, in {\it Proceedings of the Workshop on
Elementary-Particle Picture of the Universe}, eds. M Yoshimura, Y.
Totsuka, and K. Nakamura, KEK report, p. 49 (1987).

\bibitem{Wizard} R.L. Golden, S.P. Ahlen, J.J. Beatty, H.J. Crawford, P.J.
Lindstrom, J.F. Ormes, R.E. Streitmatter, C.R. Bower, R.M. Heinz, S.
Mufson, T.G. Guzik, J.P. Wefel, S.A. Stephens, J.H. Adams, K.E. Krombel,
A.J. Tylka, M. Simon, K.D. Mathis, P. Picozza, G. Barbiellini, G. Basini,
F. Bongiorno, M. Ricci, A. Codino, C. De Marzo, B. Managelli, P. Galeotti
, P. Spillantini, M. Bocciolini, Nuovo Cim. {\bf 105B}, 191 (1990).



\bibitem{TOF} {\it Time of Flight Mass Spectrometry}, American Chemical
Society symposium series, Washington, (1994); {\it Time of Flight Mass
Spectrometry}, Pergamon Press, (1969).


\end{thebibliography}
\end{document}